\documentclass[aps,superscriptaddress,twocolumn]{revtex4}

\def\urlprefix{}

\usepackage{graphicx,graphics,epsfig}   
\usepackage{dcolumn}    
\usepackage{bm}         
\usepackage{amsmath}    
\usepackage{verbatim}   
\usepackage{color}      
\usepackage{times,natbib}
\usepackage{amsmath,amsfonts,amssymb,graphics,graphics,color,times}
\usepackage{bm}
\usepackage{epstopdf}

\usepackage{latexsym}
\usepackage{amsmath}
\usepackage{amssymb}
\usepackage{amsfonts}
\usepackage{amsthm}
\usepackage{mathrsfs}
\usepackage{color,verbatim,graphics}
\DeclareMathAlphabet{\mathrsfs}{U}{rsfs}{m}{n}
\DeclareMathAlphabet{\mathpzc}{OT1}{pzc}{m}{it}
\DeclareMathAlphabet{\matheus}{U}{eus}{m}{n}
\DeclareMathAlphabet{\mathbbold}{U}{bbold}{m}{n}

\usepackage{verbatim}
\graphicspath{{graphics/}}

\def\r1{\textbf{r}}

\newcommand{\ba}{\begin{eqnarray}}
\newcommand{\ea}{\end{eqnarray}}
\newcommand{\ban}{\begin{eqnarray*}}
\newcommand{\ean}{\end{eqnarray*}}
\newcommand{\be}{\begin{equation}}
\newcommand{\ee}{\end{equation}}

\newcommand{\ket}[1]{\mathinner{|#1\rangle}}

\begin{document}

\title{Quantum modulation of a coherent state wavepacket with a single electron spin.}

\author{P. Androvitsaneas}
\altaffiliation{These authors contributed equally to this work}
\affiliation{School of Engineering, Cardiff University, Queen’s Building, The Parade, Cardiff, CF24 3AA, United Kingdom}
\affiliation{QETLabs, H. H. Wills Physics Laboratory, University of Bristol, Tyndall Avenue, Bristol, BS8 1TL, United Kingdom, and Department of Electrical \& Electronic Engineering, University of Bristol, BS8 1FD, UK}

\author{A. B. Young}
\altaffiliation{These authors contributed equally to this work}
\affiliation{QETLabs, H. H. Wills Physics Laboratory, University of Bristol,
Tyndall Avenue, Bristol, BS8 1TL, United Kingdom, and Department of Electrical \& Electronic Engineering, University of Bristol, BS8 1FD, UK}

\author{T. Nutz}
\affiliation{QETLabs, H. H. Wills Physics Laboratory, University of Bristol,
Tyndall Avenue, Bristol, BS8 1TL, United Kingdom, and Department of Electrical \& Electronic Engineering, University of Bristol, BS8 1FD, UK}

\author{J.M. Lennon}
\affiliation{QETLabs, H. H. Wills Physics Laboratory, University of Bristol,
Tyndall Avenue, Bristol, BS8 1TL, United Kingdom, and Department of Electrical \& Electronic Engineering, University of Bristol, BS8 1FD, UK}

\author{S. Mister}
\affiliation{QETLabs, H. H. Wills Physics Laboratory, University of Bristol,
Tyndall Avenue, Bristol, BS8 1TL, United Kingdom, and Department of Electrical \& Electronic Engineering, University of Bristol, BS8 1FD, UK}

\author{C. Schneider}
\affiliation{Institute of Physics, University of Oldenburg, D-26129 Oldenburg, Germany}



\author{M. Kamp}\affiliation{Technische Physik, Physikalisches Institut and Wilhelm Conrad R\"ontgen-Center for Complex Material Systems, Universit\"at W\"urzburg, Am Hubland, 97474 W\"urzburg, Germany}

\author{S. H\"ofling}\affiliation{Technische Physik, Physikalisches Institut and Wilhelm Conrad R\"ontgen-Center for Complex Material Systems, Universit\"at W\"urzburg, Am Hubland, 97474 W\"urzburg, Germany}

\author{D. P. S. McCutcheon}
\affiliation{QETLabs, H. H. Wills Physics Laboratory, University of Bristol,
Tyndall Avenue, Bristol, BS8 1TL, United Kingdom, and Department of Electrical \& Electronic Engineering, University of Bristol, BS8 1FD, UK}

\author{E. Harbord}
\affiliation{QETLabs, H. H. Wills Physics Laboratory, University of Bristol,
Tyndall Avenue, Bristol, BS8 1TL, United Kingdom, and Department of Electrical \& Electronic Engineering, University of Bristol, BS8 1FD, UK}

\author{J. G. Rarity}
\affiliation{QETLabs, H. H. Wills Physics Laboratory, University of Bristol,
Tyndall Avenue, Bristol, BS8 1TL, United Kingdom, and Department of Electrical \& Electronic Engineering, University of Bristol, BS8 1FD, UK}

\author{R. Oulton}
\affiliation{QETLabs, H. H. Wills Physics Laboratory, University of Bristol,
Tyndall Avenue, Bristol, BS8 1TL, United Kingdom, and Department of Electrical \& Electronic Engineering, University of Bristol, BS8 1FD, UK}

\begin{abstract}
The interaction of quantum objects lies at the heart of fundamental quantum physics and is key to a wide range of quantum information technologies. Photon-quantum-emitter interactions are among the most widely studied. Two-qubit interactions are generally simplified into two quantum objects in static well-defined states . In this work we explore a fundamentally new dynamic type of spin-photon interaction. We demonstrate modulation of a coherent narrowband wavepacket with another truly quantum object, a quantum dot with ground state spin degree of freedom. What results is a quantum modulation of the wavepacket phase (either $0$ or $\pi$ but no values in between), a new quantum state of light that cannot be described classically.
\end{abstract}

\maketitle

The resonantly scattered field (RSF) from a two-level-system has been extensively studied over the last 50 years \cite{HANBURYBROWN1956,PhysRevLett.39.691,PhysRev.130.2529,PhysRev.188.1969,Heitler_book} and it is considered a well understood phenomenon in physics that underpins a range of quantum technologies \cite{Divincenzo,kn-nat-409-46,PhysRevLett.92.127902,waks:153601}. The coherent fraction of RSF retains the bandwidth of the driving laser but becomes antibunched \cite{Heitler_book,PhysRevLett.108.093602}. Less explored is how the RSF is altered if the quantum emitter contains a ground state spin. For a  high magnetic field the system splits into spectrally distinct transitions that can be probed individually, where it has been shown that the RSF either retains the coherence of the drive field or undergoes dephasing on the same timescale as the electron spin T2* time \cite{PhysRevB.93.241302}. The RSF from a optically active spin is generally simplified into separate un-coupled two-level-systems \cite{PhysRevLett.92.127902, PhysRevB.78.085307,PhysRevLett.104.160503,Lindner:2009zr}, however, in the low magnetic field limit where the Zeeman splitting ($2\omega_B$) is less than the natural line-width of the optical transitions ($\Gamma$), the complex interplay between all overlapping transitions must be taken into account.

\begin{figure}
 \includegraphics[width = \linewidth]{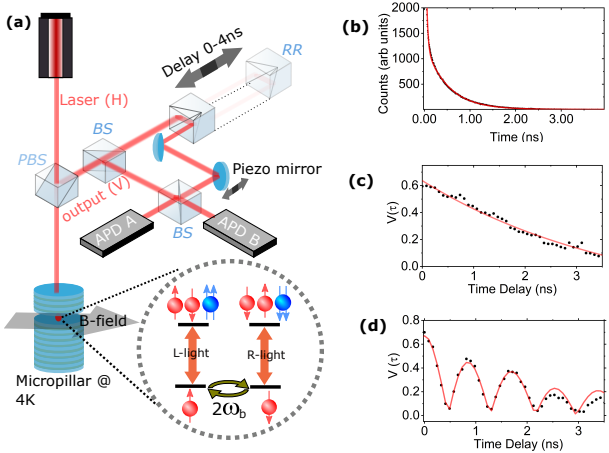}
 \caption{ Experimental setup for probing the RSF a QD spin in a pillar microcavity showing the optical selection rules for the spin in a weak in-plane (Voigt) magnetic field which induces a Larmor precession at frequency $2\omega_b$. In our experiment the drive field is a Horizontally polarised (H) single frequency laser. A polarising beam splitter (PBS) separates the H polarised pump from the V polarised signal. The output signal is split on a 50:50 beam splitter (BS) which forms the two arms of a MZI. (b) Spontaneous emission lifetime measurements obtained via pulsed resonant excitation (2ps) for QD1 where $T_1=460\pm6$ps, which corresponds to a transform limited line-width of $\Delta\omega\sim1.3\mu$eV. The initial fast decay is a result of background light from the pulsed drive laser with a timing response limited by the detector jitter ($\sim64$ps) (c),(d) Plots showing the visibility ($V(\tau)$) of the RSF as measured with the MZI in Fig.\ref{fig:main}. (a). for (c) B=0, and (d) B=108mT, for a drive field laser power $P\sim0.02P_{sat}$. The oscillations in (d) are at a frequency of $2\omega_b=590\pm10$MHz, as a result of Larmor precession and represents a sub natural line-width Zeeman splitting $\sim0.4\mu$eV}
 \label{fig:main}
\end{figure}

In this limit we reveal that the RSF does not adopt any of the previously observed behaviours, but rather the spin imparts a quantum phase modulation to the RSF. Previous RSF studies of efficient QD spin micropillar systems in high magnetic field have shown that a phase shift is imparted on the RSF \cite{Androvitsaneas:2016kx,doi:10.1021/acsphotonics.8b01380}. Since the charged QD has circularly polarised transitions (Fig.\ref{fig:main}.(a)) this phase shift can be used to rotate linearly polarized light \cite{auffeves-garnier:053823,Atature:2007zr,Young:2011uq,Sun2016,Arnold:2015bh,Androvitsaneas:2016kx,doi:10.1021/acsphotonics.8b01380}. If Horizontally polarised (H) single photons with a narrow bandwidth (weak excitation limit) scatter from the charged QD in an arbitrary pure state $\ket{\psi}_{spin}=a\ket{\uparrow}+b\ket{\downarrow}$ the cross polarised (Vertical) mode output is \cite{PhysRevB.78.085307}:

\be
\ket{\psi}_{cross}=\frac{(r_c-r_de^{i\phi_d})}{2}[a\ket{\uparrow}-b\ket{\downarrow}]\ket{V}
\label{eq1}
\ee

Here $r_c$, $r_d$ are the reflection coefficients for the empty cavity and the QD-coupled cavity respectively and $\phi_d$ is the relative phase shift induced by the QD. The term in the square brackets imparts an opposite sign to the phase for the two spin states. Both spin states rotate the polarisation to Vertical (V) but take orthogonal trajectories (clockwise/anticlockwise) around the Poincaré sphere. In the limit of zero detuning and perfect QD-cavity coupling ($\beta$-factor=1) then $r_c=r_d=1$ and $\phi=\pi$. The photon is perfectly rotated to V polarization with a global phase shift $0$ or $\pi$ depending on the spin state. The micropillar we use does not have perfect light matter interaction (i.e. $0.5<\beta$-factor$<1$), this results in $r_c$, $r_d$ being $< 1$ however this simply reduces the intensity of the RSF in the V-polarised mode, but maintains the global phase difference of $\pi$ between the two (V) RSF components, associated with $\ket{\uparrow}$ or $\ket{\downarrow}$. 

By defining the two global phase states as $\ket{+V}$ and $\ket{-V}$, one may now define the output state as:

\be
\ket{\psi}_{cross}=a\ket{\uparrow}\ket{+V}+b\ket{\downarrow}\ket{-V}
\label{eq2}
\ee

One rarely needs to consider the global phase and $\ket{\pm V}$ would not be considered as eigenstates, but an interesting situation arises when the spin evolves coherently in an in-plane (Voigt) magnetic field with a relatively slow Larmor precession period ($2\omega_b<\Gamma$). To probe these dynamics we measure the first order correlation of the RSF (Fig.\ref{fig:main}.(a)) by monitoring the visibility of the interference fringes produced over the timescale of the Larmor precession (several ns). In the Mach Zehnder interferometer (MZI) there are two optical elements that can be used to vary the path length. The first is a piezo controlled mirror that allows path length changes up to $\sim2\mu$m (with 10pm resolution). For this wavelength scale delay we assume the time delay $\sim0$ and the visibility of the observed interference fringes are given by the contrast between the maximum and minimum counts in standard silicon single photon counting avalanche photodiodes (APDs) A and B ($\sim$30\% efficient 300ps timing jitter). A separate delay stage with a retro reflector (RR) can then change the path length difference between $0-1.2$m ($\tau\sim0-4$ns). This allows an interference visibility (V) to be measured as function of time where:

\be
V(\tau)=\frac{(I^A-I^B)_{max}}{(I^A+I^B)}(\tau)\equiv|g^{(1)}(\tau)|
\ee

For a charged QD (QD1) we probe in both zero (Fig.\ref{fig:main}.(c)) and Voigt-field configurations (Fig.\ref{fig:main}.(d)) when the QD-drive laser detuning is set to zero. The measured coherence time ($T_2^*$) of the photons ($\sim3$ns) in both cases is significantly longer than the transform limit $2T_1=920\pm10$ps (Fig.\ref{fig:main}.(b)), but similar to the expected spin coherence time, as observed in \cite{PhysRevB.93.241302}. Secondly when subjected to a Voigt-magnetic field there is a pronounced oscillation in the $V(\tau)$ ($|g^{(1)}(\tau)|$) where the visibility oscillates from high to zero within an exponentially decaying envelope. This is an oscillation in the certainty with which the phase difference between two points in the wavepacket can be measured, with well-defined phase difference between half-cycles of the Larmor precession frequency, and undefined phase difference between quarter-cycles.

To illustrate why this is the case suppose at time $t=0$ the electron spin is in the state $\ket{\uparrow}$. The RSF at $t=0$ will be in the state $\ket{+V}$. The spin evolving under an in-plane field will reach at some time later ($\tau$) the superposition state $\ket{\uparrow}+\ket{\downarrow}$. At this point the RSF will evolve to be either $\ket{+V}$ or $\ket{-V}$ with an equal probability. This represents a random global phase and when the fields from $t=0$ and $t=\tau$ are interfered in an unbalanced MZI one observes a collapse in the visibility. Hence the oscillations in Fig.\ref{fig:main}.(d) can be understood as a direct mapping of the coherent Larmor precession of ($2\omega_b$) onto the phase relationship between different time points of the output RSF wavepacket. For QD1 we measure a Larmor frequency of $2\omega_b=590\pm10$MHz when $B=108$mT, corresponding to an in-plane gyromagnetic ratio $g_b=0.437$, with an inhomogeneous de-phasing time $T_2^*=2.7\pm0.1$ns. This is not the case when $B=0$ i.e. $2\omega_B=0$ or for a neutral QD with similar characteristics (see supplement).

\begin{figure}
 \includegraphics[width =\linewidth]{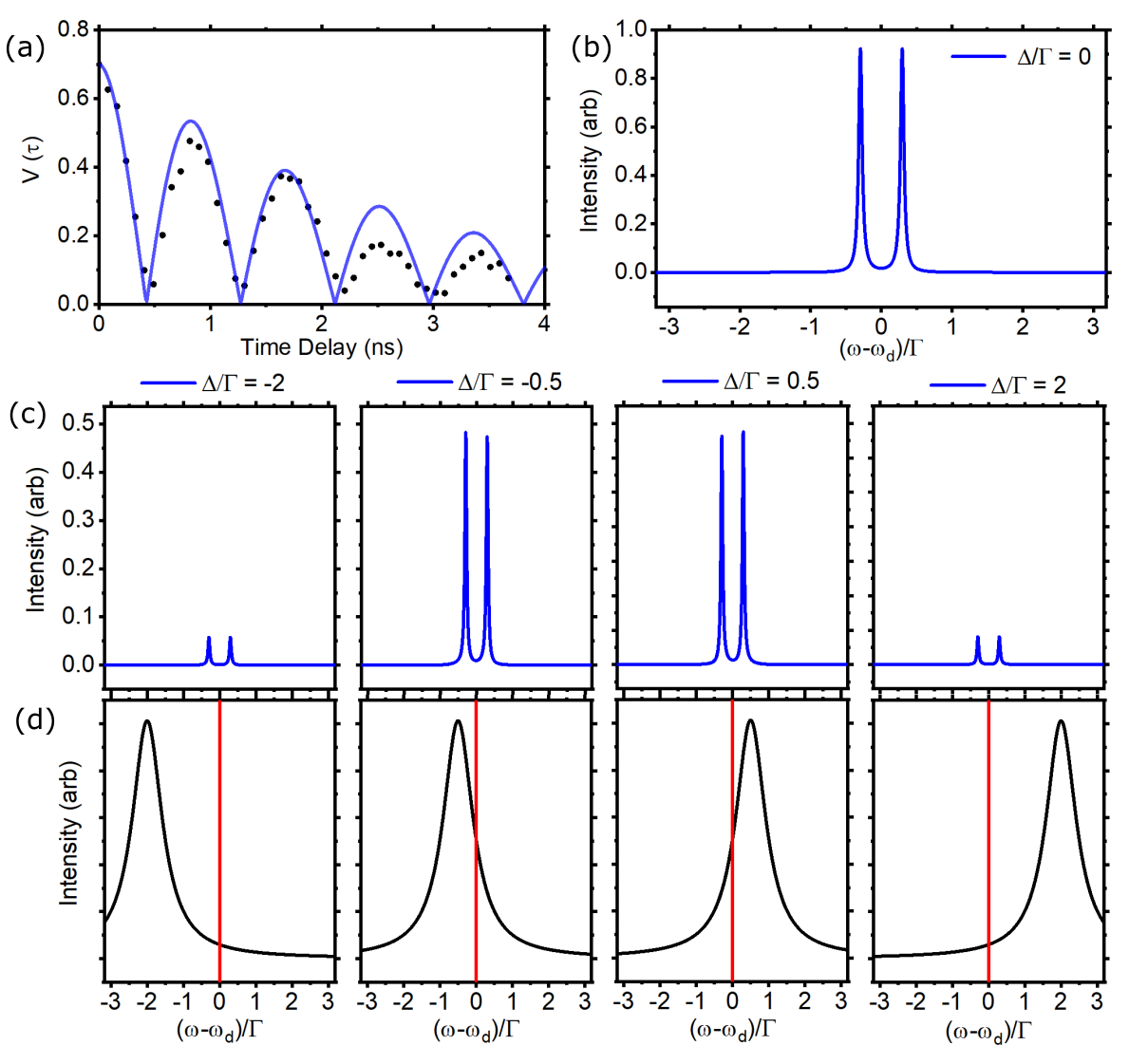}
 \caption{(a) Shows the calculated MZI visibility ($|g^{(1)}(\tau)|$) based on a numerically solved master equation (blue line) with parameters from the fits in Fig.\ref{fig:main}(d), where $|g^{(1)}(0)|$ is scaled appropriately to take into experimental imperfections. There is excellent agreement with the data (black circles). (b) Shows the corresponding spectrum of the RSF for the QD spin centred at the drive frequency of the input laser ($\omega_d$). The spectrum is plotted in units defined by the cavity enhanced spontaneous emission rate ($\Gamma$) for clarity. The QD-laser detuning is set to $\Delta=0$. The spectrum shows two peaks split by $4\omega_b\sim 0.6\Gamma$ i.e. twice the Larmor frequency each with a bandwidth $<\Gamma$. (c) Shows how the RSF spectra varies with $\Delta$, where (d) gives a corresponding visual representation of the both the QD (black curve) and laser (red line) transitions in each case. The plots in (b),(c) demonstrate that changing $\Delta$ changes the intensity of the RSF. However the spectra remains centred at the frequency of the drive laser and the position of the individual sidebands is invariant to $\Delta$. The slight asymmetry in the peak heights at moderate detuning in an artefact that arises from a pure dephasing term in the Hamiltonian which is discussed further in the supplementary information.} 
 \label{fig:control}
\end{figure}

The oscillations in the $|g^{(1)}(\tau)|$ can be compared to the numerically solved master equation for a driven QD-spin (see supplement) where we can see excellent agreement between the data and model in Fig.\ref{fig:control}.(a). This model also allows us to Fourier transform into frequency space to plot the expected spectra of the RSF. This is done via the master equation model rather than via the raw data as the oscillations have not completely disappeared after the maximum $\sim4$ns delay available in the setup in Fig.\ref{fig:main}.(a). We can see that the spectrum is split with two peaks at plus/minus the Larmor frequency (Fig.\ref{fig:control}.(b)). It is clear the RSF neither retains the coherence of the input laser or inherits the spectrum of the QD. The two peaks in the spectrum are substantially narrower than the spontaneous emission rate ($\Gamma$) and are controlled by the dephasing of the electron spin. The spectrum captures the physics controlling the RSF. There is an underlying coherent process where the input laser retains its central frequency but is modulated by the spin giving rise to sidebands in the V-polarised spectrum at $\pm$ the modulation frequency. 

These properties also ensure that the $|g^{(1)}(\tau)|$ (spectrum) is robust to spectral jitter. The inhomogeneous linewidth of the QD in Fig.\ref{fig:control}.(a) is $\sim5\mu$eV (see supplement), substantially larger than the Fourier transform limit due to charge noise imposing a time varying stark shift, altering the frequency of the transitions of the QD ($\omega$). Our fixed frequency laser ($\omega_d$) will explore a full range of detunings over the measurement and any inhomogeneities in the imparted phase shift would manifest as a rapid loss in coherence. We can calculate the effect of spectral jitter on the output RSF spectra where in Fig.\ref{fig:control}.(c) we can see the expected spectra for a range of QD-laser detunings ($\Delta=\omega-\omega_d$). It is clear that the spectrum is largely invariant to this detuning, the central frequency remains fixed at the drive laser frequency ($\omega_d$), and the splitting remains constant. The only variation is in the intensity of the RSF. For $\Delta\neq0$ we reduce the effective $\beta$-factor via detuning however, the underlying phase modulation remains fixed it is just the fraction of the input light that gets modulated changes. This provides a simple way to decouple the effects of charge and spin noise in the QD, and also confirms the phase modulation is robust to imperfections in the light-matter interaction.

\begin{figure}
 \includegraphics[width = \linewidth]{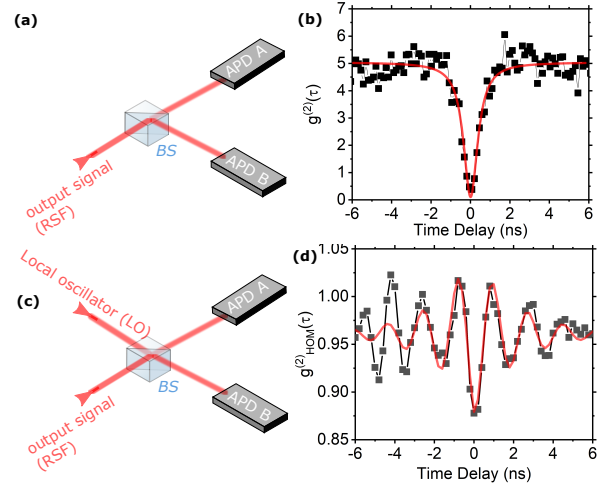}
 \caption{(a) Schematic of the HBT experiment used to measure the $g^{(2)}(\tau)$ in (b) for the case when the intensity of the LO is set to $I_{LO}=0$. Here we observe a pronounced anti-bunching feature corresponding to the exited state lifetime of the Trion state. The pronounced bunching outside of the central dip is caused by spectral jitter. (c) Schematic of the homodyne experiment used to probe the phase modulation of the RSF. The LO is derived from the same laser that is used to drive the QD. (d) $g^{(2)}_{hom}(\tau)$ for the case where the ratio of $I_{RSF}:I_{LO}$ is set to be 1:10. The y-intercept is now $\sim1$ as $I_{RSF}<<I_{LO}$. The data is fit for positive time delay only. At negative time delay the data displays an increased amplitude which is currently treated as an artefact. All data was taken at $B=108$mT with a drive field of $P\sim0.02P_{sat}$ and detected by low efficiency (few \%), but fast thin film Si APDs (MPD) with a timing jitter $\sim64$ps}
 \label{fig:g1g2comp}
\end{figure}

So far we have measured the relative phase between two points in time via a technique that is sensitive to the wave properties of the RSF.  We now explore particle-like correlations in the RSF by measuring the second-order correlation function when the RSF is combined with a local oscillator (LO). The second order correlations of the combined signal prove insightful.  Fig.\ref{fig:g1g2comp}.(b) shows the second order correlation $g^{(2)}(\tau)$ for the case when the intensity $I_{LO}=0$ in this instance the setup in Fig.\ref{fig:g1g2comp}.(a) reduces to a simple Hanbury-Brown-Twiss (HBT) interferometer and the resulting correlation shows an anti-bunching dip where $g^{(2)}(0)<0.5$ with a timescale determined by spontaneous emission rate ($\Gamma$). This antibunching of the RSF on short timescales is identical to that seen for neutral QDs \cite{PhysRevLett.108.093602} and demonstrates the single particle-like nature of the RSF. \cite{PhysRevLett.125.170402,L_pez_Carre_o_2018}.

The introduction of the LO at the other input to the BS acting as a reference phase drastically alters the output. When the LO and signal wavepackets interfere such that when their phases match, signal is directed to one detector (eg APD A), whilst when the two arms are out of phase by $\pi$, signal is directed to the opposite output (eg APD B). Thus we convert the spin dependant global phase into which path information hence project the phase of the photon onto either 0 or $\pi$ corresponding to $\ket{\uparrow}$ or $\ket{\downarrow}$. The now well-defined spin will subsequently evolve in the magnetic field, and half a precession period later will become orthogonal leading to an increase in probability of detection in APD B(A). This is evident in the second order correlation of the homodyne signal ($g^{(2)}_{hom}(\tau)$, Fig.\ref{fig:g1g2comp}.(d)) which displays a pronounced oscillation with a period of $2\omega_b=560\pm10$MHz and a coherence time $T_2^*=2.1\pm0.4$ns in line with the corresponding $|g^{(1)}(\tau)|$ data in Fig.\ref{fig:main}.(d). The fast (ns) oscillations are robust to the slow (ms) phase drift in the LO, however the resulting correlation acquired over longer timescales (hours) represents an average (random) response over both field quadratures ($\phi_{LO}=0$ and $\phi_{LO}=\pi$).

Eq.\ref{eq2} implies that for an evolving spin, only two phase values are measurable, i.e. 0 and $\pi$, with varying probability of measuring each, depending on the spin state. A classical superposition of two phase shifted waves would result in an overall phase shift of $\pi/2$. However Eq 2 implies that in quantum superposition this is not possible. To measure a $\pi/2$ phase shift in the wavepacket we now stabilise the phase of LO field to provide a well-defined reference. This allows us to measure a single time point in the wavepacket independently. 

\begin{figure}
 \includegraphics[width = \linewidth]{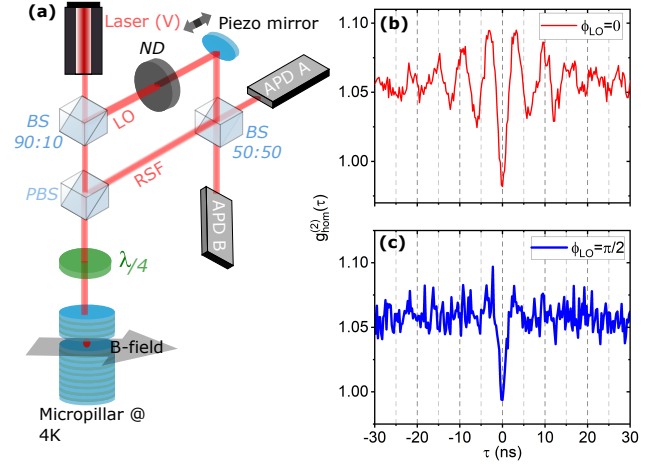}
 \caption{(a) Schematic of the stabilised homodyne interferometer. A pick off beamsplitter (BS) with a 90\% reflectivity selects non-interacting co-polarised light reflected from the micropillar that is H-polarised, this is then rotated to be V-polarised and constitutes the local oscillator (LO) arm of the interferometer. The path length difference between the LO and RSF arms of the interferometer was made equal using a (2 ps) pulsed laser to within $\sim100$ fs ($\sim30\mu$m) where a piezo mirror is responsible for fine control and is used to stabilise the path length difference. This allows us to measure intensity autocorrelations of the homodyne signal ($g^{(2)}_{hom}(\tau)$) for QD2 in two quadratures: (b) $\phi_{LO}=0$, and (c) $\phi_{LO}=\pi/2$ for a drive field power $P\sim0.1P_{sat}$ and $B=86$mT.}
 \label{fig:hom}
\end{figure}

For phase stable homodyne measurements we use QD2. This specific charged QD has a lower in-plane gyromagnetic ratio ($g\sim0.15$) and a longer $T_2^*\sim13$ns. As such we can measure correlations using standard Si-APDs (30\% efficient) with a timing jitter $\sim300$ps improving detected counts by an order of magnitude. This allows us to obtain sufficient count rate to implement a phase locked interferometer on a ms timescale. The local oscillator was derived from the co-polarised (H) channel from the setup in Fig.\ref{fig:hom}.(a) and is comprised of light that has not interacted with the QD i.e mismatched to the cavity mode and has reflected directly from the top of the micro-pillar. This minimises any correlations in the LO as a result of the QD and makes an undetectable contribution. The LO is attenuated by a neutral density (ND) such that $I_{LO}\sim10I_{RSF}$ and rotated to be V-polarised. To implement the phase lock a small amount of LO ($\delta_{LO}$) is introduced into the RSF arm using a quarter waveplate ($\sim1^o$ rotation) (see Fig.\ref{fig:hom}.(a)) such that the ratio of intensities of $I_{\delta_{LO}}:I_{RSF}$ is approximately 1:1 in the cross polarised (V) output. The $\delta_{LO}$ component then provides an interference signal when combined with the LO at the final BS that is monitored on a 15ms timescale and is used to feedback to a piezo actuated mirror to fix the phase difference between the two arms of the interferometer. By choosing different points of the interference fringe we can change the relative phase between the RSF arm and the LO arm and thus change the quadrature that we measure. This allows the whole setup to be phase stablised all the way to the sample. The noise is around 15\% about a chosen set-point as such the LO phase has the same corresponding uncertainty.


The resulting two photon correlations are presented in Fig.\ref{fig:hom} which compares the combined homodyne signal from the LO and the RSF for two cases, $\phi_{LO}=0$ and $\phi_{LO}=\pi/2$ using 256ps time bins. In Fig.\ref{fig:hom}.(b) we see that for $\phi_{LO}=0$ we observe characteristic decaying oscillations in the intensity autocorrelations of the homodyne signal ($g^{(2)}_{hom}(\tau)$), with $2\omega_b=159\pm5$MHz and $T_2^*=12.5\pm1.5$ns which correspond with the equivalent $|g^{(1)}(\tau)|$ (see supplement). In contrast for $\phi_{LO}=\pi/2$ the oscillations disappear (Fig.\ref{fig:hom}.(c)), demonstrating that at no point can the phase be defined as $\pi/2$, it can only take values of 0 or $\pi$ ($\ket{\uparrow}$ or $\ket{\downarrow}$). This contrasts sharply with a classical phase modulation that one would obtain from an electro-optic modulator \cite{Nisbet_Jones_2011,Nisbet_Jones_2013,Specht2009} which would be a continuous function of all possible phase values.

Based on our observations some interesting conclusions may be reached. Firstly, as with neutral QDs a wavepacket is scattered that is antibunched on short timescales, thus proving the single particle nature of the scattering event. Additionally Eq.\ref{eq2} reveals that the spin entirely governs the phase modulation of the wavepacket. The light-matter interaction is only sensitive to spin orientation along the optical axis, therefore the light can only detect $\ket{\uparrow}$ or $\ket{\downarrow}$. Thus only the two values 0, $\pi$ can be imparted on the wavefunction as shown in Fig.\ref{fig:hom}. Another implication is that the spin and the RSF wavepacket must be entangled. Eq.\ref{eq2} clearly predicts entanglement when the spin is in a superposition state, with results from Fig.\ref{fig:hom} implying that the entanglement persists for the timescales (ns) in the experiments.


We conclude therefore that we have demonstrated a completely new type of two-qubit interaction, a modulation of a coherent photon wavepacket with another quantum object, a spin, resulting in a new quantum state of light that cannot be described classically. This is a new class of resonant scattering with important implications for quantum information applications. From a practical point of view, while the interaction is deterministic \cite{doi:10.1021/acsphotonics.8b01380}, the bandwidth of the scattered field is several orders of magnitude narrower than that of the emitter. This along with the robustness to spectral jitter expands the scope for interfacing matter and light qubits of very different bandwidths and even wavelengths. A coherently evolving deterministic photon-spin interaction may enable easier distributed entanglement protocols, quantum switches, memories and repeaters, all mediated not by traditional single photon sources, but a narrow-band photonic "quantum bus" from a single-frequency laser. These principles are likely to apply to a wide range of spin-qubit systems where an optical or microwave interface is used, including diamond colour centres, 2D quantum emitters, superconducting qubits or atomic systems, with appropriate quantum bus light sources.

\begin{acknowledgements}
The authors acknowledge helpful discussions with A.J. Bennett. This work was funded by the Future Emerging Technologies (FET)-Open FP7-284743 [project Spin Photon Angular Momentum Transfer for Quantum Enabled Technologies (SPANGL4Q)] and the German Ministry of Education and research (BMBF) and Engineering and Physical Sciences Research Council (EPSRC) (EP/M024156/1, EP/N003381/1 and EP/M024458/1). C.S. gratefully acknowledges funding by the German Research Foundation (DPG) within the project PR1749/ 1.1. PA acknowledge financial support provided by EPSRC via Grant No. EP/T001062/1. J.M.L. and S.M. were supported by the Bristol Quantum Engineering Centre for Doctoral Training, EPSRC grant EP/L015730/1
\end{acknowledgements}
\end{document}